\documentclass[preprint,12pt]{elsarticle}

\usepackage{amssymb}
\usepackage{epsfig}
\usepackage{epsf}
\usepackage{booktabs}
\usepackage{amssymb}
\usepackage{multirow}
\usepackage{mathrsfs}
\usepackage{longtable}
\usepackage{lscape}
\usepackage{tabularx}
\usepackage{subfigure}

\newdefinition{definition}{Definition}
\newdefinition{example}{Example}
\newdefinition{remark}{Remark}

\journal{Journal of Theoretical Biology}

\begin{document}

\begin{frontmatter}

\title{Generalized prisoner's dilemma}

\author[SWU]{Xinyang Deng}
\author[VUMC]{Qi Liu}
\author[SWU,VU]{Yong Deng\corref{COR}}
\ead{ydeng@swu.edu.cn, prof.deng@hotmail.com}
\cortext[COR]{Corresponding author: Yong Deng, School of Computer and Information Science, Southwest University, Chongqing, 400715, China}

\address[SWU]{School of Computer and Information Science, Southwest University, Chongqing, 400715, China}
\address[VUMC]{Department of Biomedical Informatics, Medical Center, Vanderbilt University, Nashville, TN, 37235, USA}
\address[VU]{School of Engineering, Vanderbilt University, Nashville, TN, 37235, USA}

\begin{abstract}
Prisoner's dilemma has been heavily studied. In classical model, each player chooses to either "Cooperate" or "Defect". In this paper, we generalize the prisoner's dilemma with a new alternative which is neither defect or cooperation. The classical model is the special case under the condition that the third state is not taken into consideration.
\end{abstract}


\end{frontmatter}



\section{Introduction}
Game theory is exploited by von Neumann and Morgenstern \cite{von2007theory} and developed by many researchers, such as Nash \cite{nash1950equilibrium} and Shapley \cite{shapley1952value}. This theory is extensively used in many fields, especially economics \cite{dixit1999games,fudenberg1998theory} and biology \cite{smith1982evolution,nowak2006evolutionary}. The current study is mainly focus on how to promote the cooperation in a game \cite{marcoux2013network,yonenoh2014selection}. Game theory provides a mathematical framework to explain and address the interactive decision situations where the aims, goals and preferences of the participating agents are potentially in conflict \citep{chen2011combined,butnariu2008shapley,li2010collective}. A strategic game contains three parts: set of players, set of strategies for each player and payoff for each strategy combination, respectively. A solution to a game is a certain combination of strategies. This solution which is self enforcing and no player can gain by unilaterally deviating from it, is said to be a Nash equilibrium \cite{nash1950equilibrium}.

Two-person non-constant sum game is a kind of widely addressed game, such as iterated prisoner's dilemma game. In this game, player 1 has a finite strategy set $S_1$ including $p$ strategies. Player 2 has a finite strategy set $S_2$ including $q$ strategies. The payoffs of player 1 and player 2 are determined by functions $u_1(s_1, s_2)$ and $u_2(s_1, s_2)$, respectively, where $s_1 \in S_1$ and $s_2 \in S_2$. A combination of each players' strategy $(s_1^{*}, s_2^{*})$ is a Nash equilibrium of this two-person non-constant sum game if
\begin{equation}\label{Eq1}
u_1 (s_1^* ,s_2^* ) \ge u_1 (s_1 ,s_2^* )\quad \forall s_1  \in S_1
\end{equation}
\begin{equation}\label{Eq2}
u_2 (s_1^* ,s_2^* ) \ge u_2 (s_1^* ,s_2 )\quad \forall s_2  \in S_2
\end{equation}

\section{Prisoner's dilemma}
This game, the ``prisoner's dilemma" \cite{poundstone2011prisoner}, is a classic model of two-person non-constant sum game. It described this situation. Two criminal suspects are arrested and imprisoned. Each prisoner is in solitary confinement with no means of speaking to or exchanging messages with the other. The police admit they don't have enough evidence to convict the pair on the principal charge. They plan to sentence both to a year in prison on a lesser charge. Simultaneously, the police offer each prisoner a Faustian bargain. Each prisoner is given the opportunity either to betray the other or to cooperate with the other. Here's how it goes:
\begin{enumerate}[(i)]
  \item If A and B both betray the other, each of them serves 4 years in prison;
  \item If A betrays B but B cooperates with A, A will be set free and B will serve 5 years in prison (and vice versa);
  \item If A and B both betray the other, both of them will only serve 1 year in prison (on the lesser charge).
\end{enumerate}

\begin{figure}[htbp]
\begin{center}
\psfig{file=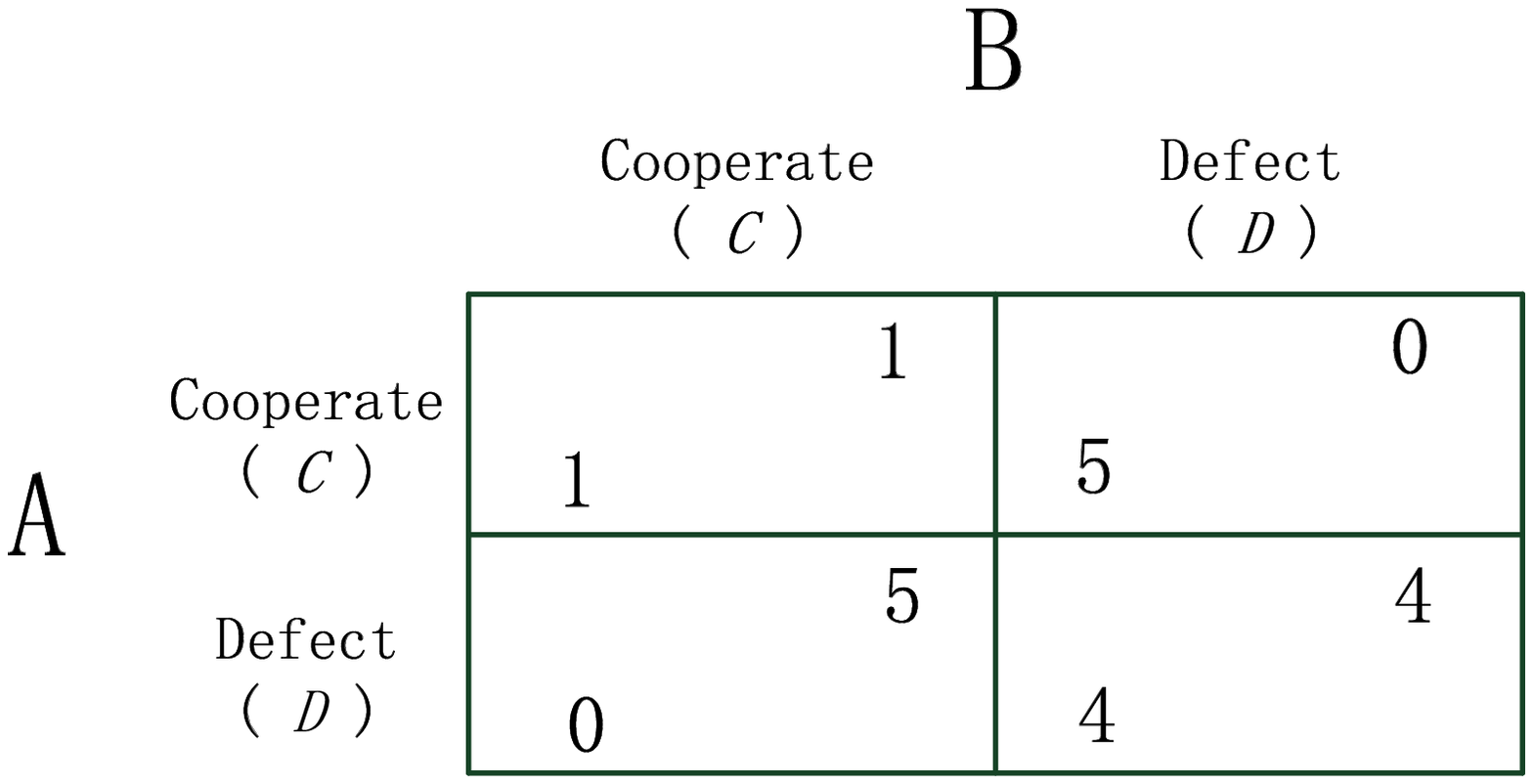,scale=0.6,angle=-90}
\caption{Prisoner's dilemma}\label{PDgame}
\end{center}
\end{figure}

The game can be shown as Figure \ref{PDgame}. Based on Eq.(\ref{Eq1}) and (\ref{Eq2}), its Nash equilibrium is ($D$, $D$).

\section{Proposed generalized prisoner's dilemma}
\begin{figure}[htbp]
\begin{center}
\psfig{file=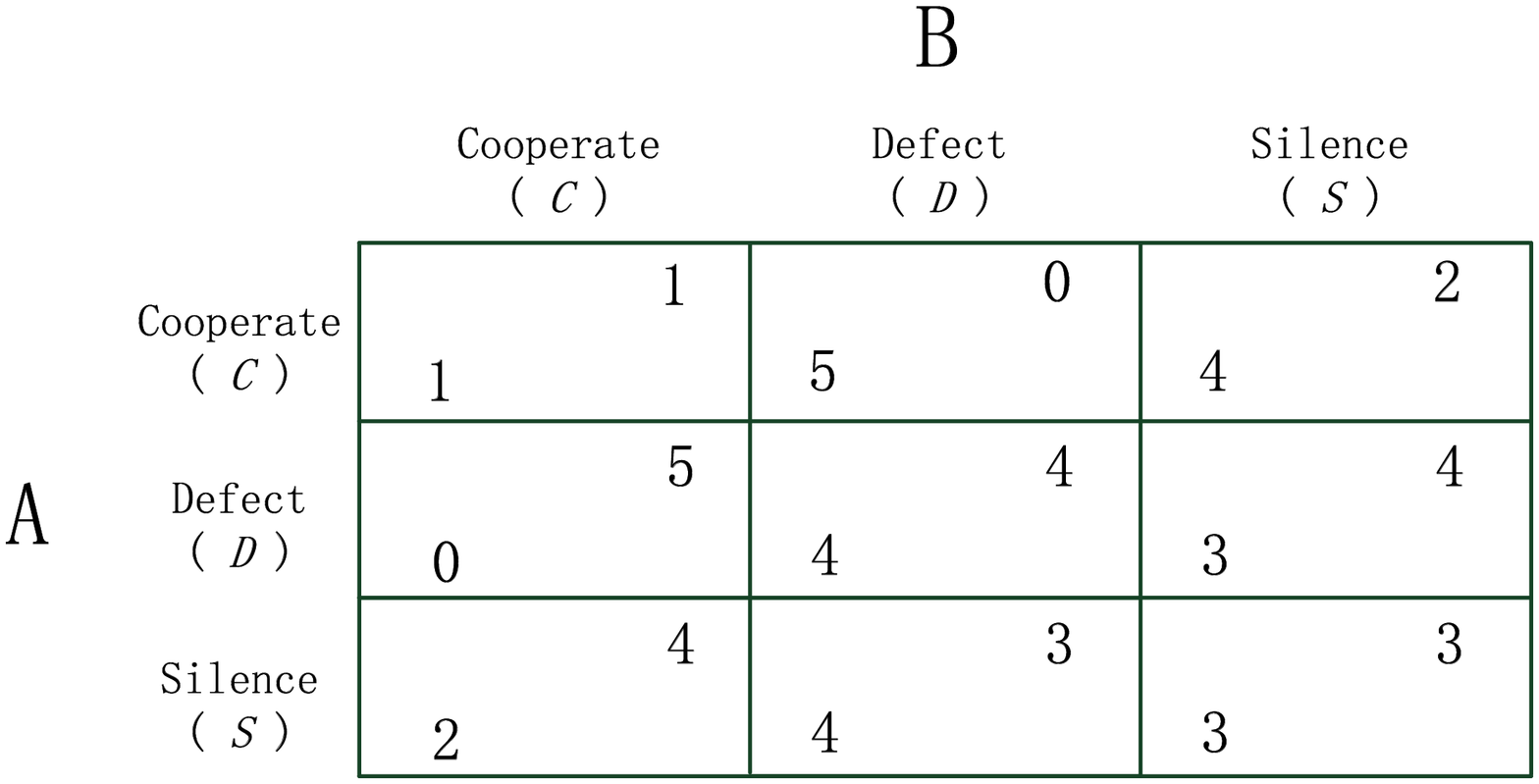,scale=0.6,angle=-90}
\caption{generalized prisoner's dilemma}\label{GPDgame}
\end{center}
\end{figure}

In the classical prisoner's dilemma, each prisoner has two strategies, namely cooperate ($C$) and defect ($D$). However, the prisoner may have another choice that is to silence ($S$). To be silence means that neither C nor D has been chosen by the prisoners. If a prisoner's strategy is S, it means his attitude is ambiguous. Wave-particle duality is a well known theory that proposes that all matter exhibits the properties of not only particles, which have mass, but also waves, which transfer energy. Similar to this theory, the duality also exists in  prisoner's dilemma. In the classical prisoner's dilemma, each prisoner has two strategies, namely cooperate ($C$) and defect ($D$). However, the prisoner may have another choice that is to silence ($S$). To be silence means that neither C nor D has been chosen by the prisoners. If a prisoner's strategy is S, it means his attitude is ambiguous. Similar with the light which has wave-particle dualism, strategy S implies the the prisoner can be either $C$ or $D$, but the police can not distinguish. In this case, $S$ is a special state which is a mixture of $C$ and $D$. In this situation, the strategy set of each prisoner is $\{C, D, S\}$. An example of this case is shown as Figure \ref{GPDgame}.

In the classical prisoner's dilemma model, the third state $S$ is not taken into consideration. The proposed model is a generalization of classical prisoner's dilemma which can be reduced to the classical prisoner's dilemma when the strategy $S$ can be distinguished as $C$ or $D$.

\section{Conclusion}
In this paper, a generalized prisoner's dilemma model is proposed. Within the proposed model, apart from ``Cooperate" and ``Defect", the third state ``Silence" has been take into consideration. The classical model is the special case of proposed generalized prisoner's dilemma model under the condition that the third state is not taken into consideration.

\section*{Acknowledgement}
The work is partially supported by National Natural Science Foundation of China (Grant nos. 61174022 and 71271061), National High Technology Research and Development Program of China (863 Program) (Grant no. 2013AA013801), R \& D Program of China (2012BAH07B01).





\bibliographystyle{elsarticle-num}
\bibliography{reference}

\begin{thebibliography}{10}
\expandafter\ifx\csname url\endcsname\relax
  \def\url#1{\texttt{#1}}\fi
\expandafter\ifx\csname urlprefix\endcsname\relax\def\urlprefix{URL }\fi
\expandafter\ifx\csname href\endcsname\relax
  \def\href#1#2{#2} \def\path#1{#1}\fi

\bibitem{von2007theory}
J.~von Neumann, O.~Morgenstern, Theory of Games and Economic Behavior (60th
  Anniversary Commemorative Edition), Princeton university press, 2007.

\bibitem{nash1950equilibrium}
J.~F. Nash, et~al., Equilibrium points in n-person games, Proceedings of the
  national academy of sciences 36~(1) (1950) 48--49.

\bibitem{shapley1952value}
L.~S. Shapley, A value for n-person games, Tech. rep., DTIC Document (1952).

\bibitem{dixit1999games}
A.~K. Dixit, S.~Skeath, D.~Reiley, Games of strategy, Norton New York, 1999.

\bibitem{fudenberg1998theory}
D.~Fudenberg, The theory of learning in games, Vol.~2, MIT press, 1998.

\bibitem{smith1982evolution}
J.~M. Smith, Evolution and the Theory of Games, Cambridge university press,
  1982.

\bibitem{nowak2006evolutionary}
M.~A. Nowak, Evolutionary dynamics: exploring the equations of life, Harvard
  University Press, 2006.

\bibitem{marcoux2013network}
M.~Marcoux, D.~Lusseau, Network modularity promotes cooperation, Journal of
  theoretical biology 324 (2013) 103--108.

\bibitem{yonenoh2014selection}
H.~Yonenoh, E.~Akiyama, Selection of opponents in the prisoner's dilemma in
  dynamic networks - an experimental approach, Journal of theoretical biology
  351 (2014) http://dx.doi.org/10.1016/j.jtbi.2014.02.006.

\bibitem{chen2011combined}
B.~Chen, B.~Zhang, W.~Zhu, Combined trust model based on evidence theory in
  iterated prisoner's dilemma game, International Journal of Systems Science
  42~(1) (2011) 63--80.

\bibitem{butnariu2008shapley}
D.~Butnariu, T.~Kroupa, Shapley mappings and the cumulative value for n-person
  games with fuzzy coalitions, European Journal of Operational Research 186~(1)
  (2008) 288--299.

\bibitem{li2010collective}
J.~Li, G.~Kendall, Collective behavior and kin selection in evolutionary
  iterated prisoner's dilemma, Journal of Multiple-Valued Logic and Soft
  Computing 16~(6) (2010) 509--525.

\bibitem{poundstone2011prisoner}
W.~Poundstone, Prisoner's dilemma, Random House LLC, 2011.

\end{thebibliography}







\end{document}